\documentclass[twocolumn,showpacs,superscriptaddress,floatfix,preprintnumbers,amssymb,amsmath]{revtex4}

\usepackage{graphicx}% Include figure files
\usepackage{dcolumn}% Align table columns on decimal point
\usepackage{bm}% bold math
\usepackage[latin1]{inputenc}
\usepackage[mathscr]{eucal}
\usepackage{epsfig}
\usepackage{rotating}

\begin{document}

\title{Nanoampere pumping of Cooper pairs}

\author{Juha J. Vartiainen}
\affiliation{Low Temperature Laboratory, Helsinki University of
Technology, P.O. Box 3500, 02015 TKK, Finland}
\author{Mikko M\"{o}tt\"{o}nen}
\affiliation{Low Temperature Laboratory, Helsinki University of
Technology, P.O. Box 3500, 02015 TKK, Finland}
\affiliation{Laboratory of Physics, Helsinki University of Technology,
P. O. Box 4100, 02015 TKK, Finland}
\author{Jukka P. Pekola}
\affiliation{Low Temperature Laboratory, Helsinki University of
Technology, P.O. Box 3500, 02015 TKK, Finland}
\author{Antti Kemppinen}
\affiliation{Centre for Metrology and Accreditation (MIKES), Electricity Group, P.O. Box 9, 02151 ESPOO, Finland}

\begin{abstract}
We have employed a tunable Cooper-pair transistor, the sluice,
with radio frequency control to pump current over a resistive circuit. We find that the charge
transferred per pumping cycle can be controlled with the resolution of
a single Cooper-pair up to hundreds of pairs. The achieved nanoampere
current features more than an order of magnitude improvement over the previously
reported results and it is close to the theoretical maximum value for the measured sample.
\end{abstract}


\maketitle The discreteness of
electron charge together with good controllability of high-frequency
signals renders a tunnel-junction based electron pump a potential
quantum standard of electric current~\cite{zimmerman}. The distinctive
feature of the electron pump is to transfer a known
multiple $n$ of the elementary charge $e$ at a fixed frequency $f$
resulting in an average current
\begin{equation}
\label{virta}
I_{\rm pump}=nef.
\end{equation}
The most precise electron pump has been demonstrated
with a relative accuracy~$10^{-8}$ using seven
normal-state tunnel junctions in series~\cite{Keller1996}.
In this design, however, the pumping is limited by the $RC$ time constant determined by the fixed tunneling resistance
and capacitance of the island. Hence, the highest achievable current is on picoampere level which is well
below a nanoampere desired for the so-called quantum triangle experiment~\cite{Likharev1985}.
The first attempts~\cite{geerligs,7b,Toppari} to build a superconducting Cooper-pair pump by replacing
the tunnel junctions by Josephson junctions were also $RC$ time limited and, in addition, suffered from considerable leakage current.

In microstructures, the tunable tunneling barriers or charge confinement
can be arranged with the help of superconducting quantum interference devices~\cite{niskanenpumppu} (SQUIDs),
mechanical motion~\cite{isacsson:277002}, or by engineering spatial electric potentials~\cite{Si,saw}.
To date, the only class of single charge pumps which generate nanoampere currents,
albeit not yet reaching the metrological accuracy,
are based on electrons carried by surface-acoustic waves~\cite{saw}. On the other hand,
the maximal reported currents obtained with SQUID based pumps are tens of picoamperes
although theoretically higher currents should be possible.

In this Letter, we report measurements on a Cooper-pair sluice involving two controllable SQUIDs and one gate resulting in current of 1~nA.
Advantages in the layout design, fabrication of
a homogenous junction set, and improvements in the control pulse sequences allow us to pump several hundreds of Cooper-pairs per cycle which is more than
a decade higher than in previous experiments. Thus a pumped current of a nanoampere level can be reached with frequencies of a few tens of MHz.
Ultimately, the critical currents~$I_c$ of the Josephson junctions of the pump limit the highest current possible to pass through the structure.
Taking into account that only part of the duty cycle transfers charge through a particular junction
we are close but do not meet this limit with $I_c\sim20~\textrm{nA}$ in our sample.

\begin{figure}[h]
    \begin{center}
    \includegraphics[width=.45\textwidth]{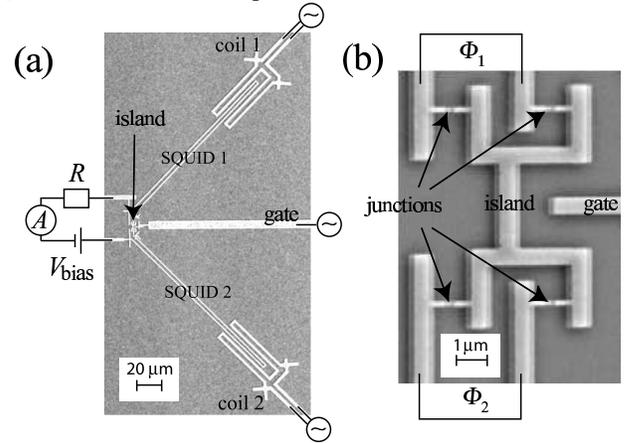}
    \end{center}
   \caption{\label{fig:sample} Scanning electron micrograph of the sample with  a sketch of simplified measurement setup.
   (a) Overall sample layout showing on-chip coils and SQUIDs separating the island. Here the resistance in series with the pump is $R=5~\textrm{k}\Omega$.
   The additional cross-shaped structures serve to absorb the stitching errors in the lithography.
   (b) Magnified view of the island with four Josephson junctions.}
\end{figure}

The measured sample consists of a $\mu$m-scale island linked to the
leads by two SQUIDs, see Fig.~\ref{fig:sample}. Each of the SQUIDs consists of two  AlOx tunnel barriers of lateral size
60~nm~$\times$~100~nm fabricated by standard electron beam lithography and two-angle evaporation into an all-aluminium device on oxidized silicon wafer.
The detailed description of the operational principle and the measurement set-up of the sluice is
presented in Refs.~\onlinecite{niskanenpumppu} and~\onlinecite{Niskanen2005}.
All the measurements were performed at sub-200~mK temperatures in a ${}^3$He-${}^4$He
dilution refrigerator.

The normal state resistance of the device is $R_{\rm N}=16.1$~k$\Omega$. This corresponds to critical current $I_{\rm c}=19.5$~nA
and Josephson energy $E_{\rm J}/k_{\rm B}=460$~mK for a single junction according to Ambegaokar-Baratoff formula.
We measured the Coulomb-blockade peak
in the differential conductance~\cite{cbt} at 4.2~K yielding the total capacitance $C_{\Sigma}=2.3$~fF
and the corresponding charging energy $E_{\rm C}/k_{\rm B}=400$~mK of the island.

\begin{figure}[h]
    \begin{center}
    \includegraphics[width=.45\textwidth]{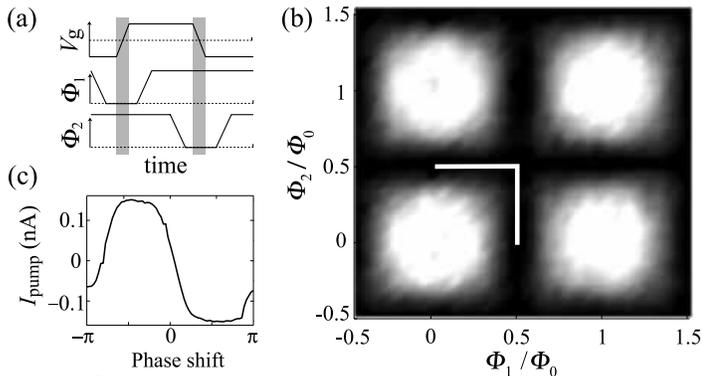}
    \end{center}
   \caption{\label{fig:control} (a) Employed pulse sequence to control the gate voltage and the fluxes. The gate sweeps
transferring charge through the SQUIDs are emphasized by gray columns.
   (b) Measured zero bias differential conductance
   as a function of static fluxes through SQUIDs~1 and~2.
   The black color marks the minimum and white the maximum conductance which is four decades higher than the minimum. Here the data
   is averaged over all possible gate voltages. The path
   traversed in this flux plane during the pumping cycle is marked by white lines.
   (c) Measured pumped current as a function of the phase shift of the gate pulse with respect to the flux pulses.}
\end{figure}

We drive the pump by sending flux pulses to the SQUIDS using the on-chip coils
and manipulating the island charge by a voltage on the gate synchronously, see Fig.~\ref{fig:control}(a).
The pumping period begins with a flux ramp $\Phi_1: \Phi_0/2 \rightarrow 0$ which opens the SQUID~1 and a consequent gate ramp transferring
desired amount of charge to the island. In the latter part of the cycle, we close SQUID~1
and open SQUID~2, after which the gate is ramped to its initial level. This channels the island charge
through SQUID~2 which is thereafter closed again.
The signal involves flat sections between the ramps allowing relaxation of possible transients
developing due to the finite band width of the signal input lines. The pulse
parameters can be estimated from DC-measurements and fine tuned for each driving frequency as discussed below.

Figure~\ref{fig:control}(b) illustrates how the magnetic fluxes control the zero bias differential conductance and hence
the tunneling rates through the SQUIDs. Let us denote the mutual inductance from coil 1 (2) to SQUID~1 (2) by~$M_{11}$ ($M_{22}$) and the cross
coupling from coil~1 (2) to SQUID~2 (1) by~$M_{12}$ ($M_{21}$). The measured inductance matrix for the reported sample is
\begin{equation}
\label{eq:indu}
M=\left(
  \begin{array}{cc}
   6.5 & 0.06 \\
    0.12 & 6.6 \\
  \end{array}
\right)\textrm{pH}.
\end{equation}
The critical current of the on-chip coils and inductances~$M_{11}$ and~$M_{22}$ allow one to sweep over at least five flux quanta
$\Phi_0\approx 2.07 \times 10^{-15}$ Wb through both SQUIDs. Currents above~$\sim\!\!1~\textrm{mA}$ in the 2~$\mu$m wide and 100~nm
thick superconducting coils drive them into a resistive state resulting in local heating.
However, only flux values from 0 to $\Phi_0/2$ are needed in the pumping experiment corresponding to 0 -- 0.15~mA currents in the coils.
Since the parasitic inductances~$M_{12}$ and~$M_{21}$ were negligible compared with other error sources, we
do not need additional current pulses to compensate for the cross coupling~\cite{Niskanen2005}.

In the ideal operation of the sluice, one of the SQUIDs is always closed. However, non-zero residual Josephson energies of the SQUIDs introduce
leakage and pumping errors~\cite{niskanenpumppu,Niskanen2005}. We minimize the leakage current
by sweeping the flux pulse offsets and their relative phase shift while driving the rf-pulses on coils.
Here we apply a constant gate $V_{\rm g}=0$ and bias voltage $V_{\rm bias}=0.1$ mV.

From the $e$-periodic modulation of the $IV$-characteristics as a function of gate voltage, we extract the gate capacitance $C_{\rm g}$ which is about $0.3$~fF.
We let the gate pulse to be symmetric with respect to zero voltage to avoid unintentional bias over the sample and denote its amplitude by $V_{\rm g}^{\rm max}$.
Figure~\ref{fig:control}(c) illustrates the pumped current as a function of the phase shift of the gate pulse with respect to
the flux pulses. For each operation frequency, we swept over the full range of phase shifts and selected the one which yields
the largest pumped current. Due to the flat sections of this curve, the pumped current is insensitive to the phase shift at the selected point.
Note that a phase shift of $\pi$ results in pumping in the opposite direction.

\begin{figure}[h]
    \begin{center}
    \includegraphics[width=.45\textwidth]{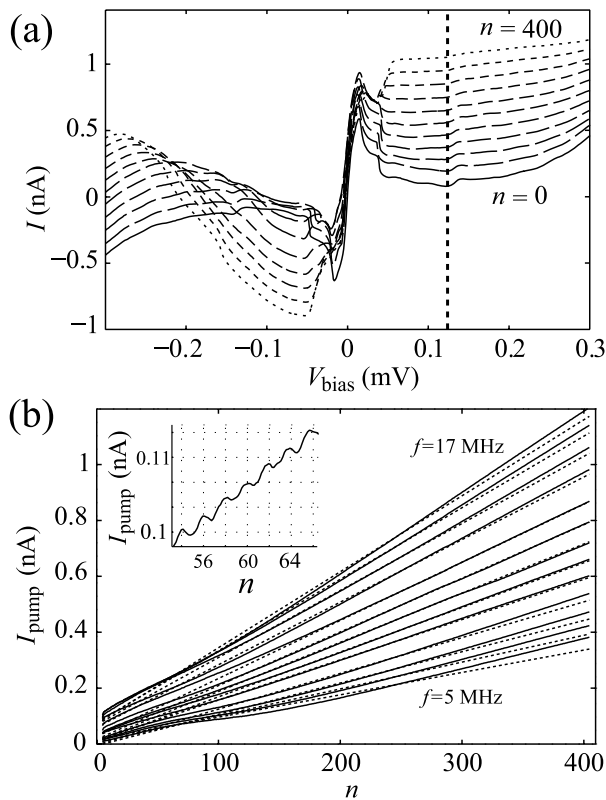}
    \end{center}
    \caption{\label{fig:pumppaus} (a) $IV$-characteristics for pumping at 15~MHz
    with gate induced charge ranging form $n=0$ to $n=400$ in uniform intervals.
    Dashed vertical line shows the selected operational point in voltage bias.
    (b) Pumped current as a function of gate-induced charge for selected frequencies.
    The dotted $I=nef+I_{\rm leak}$ lines are forced match the measurement data at $n=250$ with $I_{\rm leak}$ as a fitting parameter.
    The inset shows the step-like behavior of the pumping.
    }
\end{figure}

We study the pumped current as a function of the gate-induced charge~$n=2V_{\rm g}^{\rm max}C_{\rm g}/e$,
operational frequency $f$, and bias voltage $V_{\rm bias}$, see Fig.~\ref{fig:pumppaus}.
Figure~\ref{fig:pumppaus}(a) shows that the pump generates an approximately constant current of desired magnitude in a wide
region of positive $V_{\rm bias}$. The pumping is sensitive to the operational point in the steep regions of the $IV$-curve.
In contrast, the operational points where the $IV$-curve achieves a local minimum are stable since the current noise due to voltage fluctuations
vanishes up to linear order. Moreover, this point yields a local minimum of leakage.

In Fig.~\ref{fig:pumppaus}(b), we focus on the  $n$ dependence of the pumped current. We note that the measured slope of the pumped current
is close to the theoretical value, though some discrepancy still remains.
Although we minimize the current leakage with available control parameters to find the optimal $V_{\rm bias}$, leakage of tens of picoamperes persists,
see the value of $I$ at $n=0$ in Fig.~\ref{fig:pumppaus}(b).
The inset of Fig.~\ref{fig:pumppaus}(b) shows steps of height~$2ef$ in the pumped current,
which are appearing in two-electron intervals in $n$ due to the symmetric gate pulse. The possibility to count these steps from zero to large $n$ gives
a calibration for the gate pulse amplitude.
The fluxes through the SQUIDs control the tunneling rates of the Cooper pairs but not those of the quasiparticles, i.e., unpaired electrons.
Hence, the number of electrons pumped per cycle must be even. However, we record data which is an average over
even and odd number of excess electrons on the island
since the tunneling rates of the quasiparticles ($\sim0.1~\mu \textrm{s}^{-1}$ in Ref.~\onlinecite{Naaman}) are much faster than our measurement
time $\sim0.1~$s per data point.

The observed results can be modeled using coherent~\cite{Niskanen2005} or incoherent~\cite{ingold,lotkhov2004} theories. However,
the measured sample parameters fall into a regime where neither of them is strictly valid. If the pump was embedded in a highly dissipative environment
in form of nearby on-chip resistors, which might also help to suppress the leakage, the operation of the device could possibly
be explained in the approximation of discrete tunneling events~\cite{Lotkhov}. In our design however,
the resistors cannot be placed near the junctions because the SQUID loops are relatively large.

To further reduce the residual $E_\textrm{J}$ and leakage current, fabrication of identical junctions in each SQUID is required.
Our experiments indicate (data not shown) that a SQUID with three junctions~\cite{niskanenpumppu} helps in
solving this inhomogeneity issue. Another approach is to replace the SQUIDs by more
sophisticated topologies with several junctions in parallel and in series~\cite{Cholascinski}.
However, these structures require control over all the fluxes through
the various loops and the gate charge of all the islands, which is an experimental challenge.

In conclusion, we have demonstrated synchronized charge transfer on 1~nA level in a Cooper-pair sluice, still maintaining the step-like structure in the
pumped current as a function of the gate amplitude. Besides the metrological application~\cite{zimmerman}, the large current opens a
possibility to use the sluice to measure the Berry phase in a superconducting circuit~\cite{Mottonen2006}.

We acknowledge Academy of Finland, Finnish Cultural Foundation, and V\"ais\"al\"a foundation for financial support. We thank Kurt Baarman
for help in building the measurement set-up.


\begin{thebibliography}{18}
\expandafter\ifx\csname
natexlab\endcsname\relax\def\natexlab#1{#1}\fi
\expandafter\ifx\csname bibnamefont\endcsname\relax
  \def\bibnamefont#1{#1}\fi
\expandafter\ifx\csname bibfnamefont\endcsname\relax
  \def\bibfnamefont#1{#1}\fi
\expandafter\ifx\csname citenamefont\endcsname\relax
  \def\citenamefont#1{#1}\fi
\expandafter\ifx\csname url\endcsname\relax
  \def\url#1{\texttt{#1}}\fi
\expandafter\ifx\csname urlprefix\endcsname\relax\def\urlprefix{URL
}\fi \providecommand{\bibinfo}[2]{#2}
\providecommand{\eprint}[2][]{\url{#2}}

\bibitem[{\citenamefont{Zimmerman and Keller}(2003)}]{zimmerman}
\bibinfo{author}{\bibfnamefont{N.~M.} \bibnamefont{Zimmerman}}
  \bibnamefont{and} \bibinfo{author}{\bibfnamefont{M.~W.}
  \bibnamefont{Keller}}, \bibinfo{journal}{Meas. Sci. Technol.}
  \textbf{\bibinfo{volume}{14}}, \bibinfo{pages}{1237} (\bibinfo{year}{2003}).

\bibitem[{\citenamefont{Keller et~al.}(1996)\citenamefont{Keller, Martinis,
  Zimmerman, and Steinbach}}]{Keller1996}
\bibinfo{author}{\bibfnamefont{M.~W.} \bibnamefont{Keller}},
  \bibinfo{author}{\bibfnamefont{J.~M.} \bibnamefont{Martinis}},
  \bibinfo{author}{\bibfnamefont{N.~M.} \bibnamefont{Zimmerman}},
  \bibnamefont{and} \bibinfo{author}{\bibfnamefont{A.~H.}
  \bibnamefont{Steinbach}}, \bibinfo{journal}{Appl. Phys. Lett.}
  \textbf{\bibinfo{volume}{69}}, \bibinfo{pages}{1804} (\bibinfo{year}{1996}).

\bibitem[{\citenamefont{Likharev and Zorin}(1985)}]{Likharev1985}
\bibinfo{author}{\bibfnamefont{K.~K.} \bibnamefont{Likharev}} \bibnamefont{and}
  \bibinfo{author}{\bibfnamefont{A.~B.} \bibnamefont{Zorin}},
  \bibinfo{journal}{J.\ Low Temp.\ Phys.} \textbf{\bibinfo{volume}{59}},
  \bibinfo{pages}{347} (\bibinfo{year}{1985}).

\bibitem[{\citenamefont{Geerligs et~al.}(1991)\citenamefont{Geerligs, Verbrugh,
  Hadley, Mooij, Pothier, Lafarge, Urbina, Esteve, and Devoret}}]{geerligs}
\bibinfo{author}{\bibfnamefont{L.~J.} \bibnamefont{Geerligs}},
  \bibinfo{author}{\bibfnamefont{S.~M.} \bibnamefont{Verbrugh}},
  \bibinfo{author}{\bibfnamefont{P.}~\bibnamefont{Hadley}},
  \bibinfo{author}{\bibfnamefont{J.~E.} \bibnamefont{Mooij}},
  \bibinfo{author}{\bibfnamefont{H.}~\bibnamefont{Pothier}},
  \bibinfo{author}{\bibfnamefont{P.}~\bibnamefont{Lafarge}},
  \bibinfo{author}{\bibfnamefont{C.}~\bibnamefont{Urbina}},
  \bibinfo{author}{\bibfnamefont{D.}~\bibnamefont{Esteve}}, \bibnamefont{and}
  \bibinfo{author}{\bibfnamefont{M.~H.} \bibnamefont{Devoret}},
  \bibinfo{journal}{Z. Phys. B: Condens. Matter} \textbf{\bibinfo{volume}{85}},
  \bibinfo{pages}{349} (\bibinfo{year}{1991}).

\bibitem[{\citenamefont{Aumentado et~al.}(2003)\citenamefont{Aumentado, Keller,
  and Martinis}}]{7b}
\bibinfo{author}{\bibfnamefont{J.}~\bibnamefont{Aumentado}},
  \bibinfo{author}{\bibfnamefont{M.~W.} \bibnamefont{Keller}},
  \bibnamefont{and} \bibinfo{author}{\bibfnamefont{J.~M.}
  \bibnamefont{Martinis}}, \bibinfo{journal}{Physica E}
  \textbf{\bibinfo{volume}{18}}, \bibinfo{pages}{37} (\bibinfo{year}{2003}).

\bibitem[{\citenamefont{Toppari et~al.}(2004)\citenamefont{Toppari, Kivioja,
  Pekola, and Savolainen}}]{Toppari}
\bibinfo{author}{\bibfnamefont{J.~J.} \bibnamefont{Toppari}},
  \bibinfo{author}{\bibfnamefont{J.~M.} \bibnamefont{Kivioja}},
  \bibinfo{author}{\bibfnamefont{J.~P.} \bibnamefont{Pekola}},
  \bibnamefont{and} \bibinfo{author}{\bibfnamefont{M.~T.}
  \bibnamefont{Savolainen}}, \bibinfo{journal}{J.\ Low Temp.\ Phys.}
  \textbf{\bibinfo{volume}{136}}, \bibinfo{pages}{1573} (\bibinfo{year}{2004}).

\bibitem[{\citenamefont{Niskanen et~al.}(2003)\citenamefont{Niskanen, Pekola,
  and Sepp\"{a}}}]{niskanenpumppu}
\bibinfo{author}{\bibfnamefont{A.~O.} \bibnamefont{Niskanen}},
  \bibinfo{author}{\bibfnamefont{J.~P.} \bibnamefont{Pekola}},
  \bibnamefont{and}
  \bibinfo{author}{\bibfnamefont{H.}~\bibnamefont{Sepp\"{a}}},
  \bibinfo{journal}{Phys. Rev. Lett.} \textbf{\bibinfo{volume}{91}},
  \bibinfo{pages}{177003} (\bibinfo{year}{2003}).

\bibitem[{\citenamefont{Isacsson et~al.}(2002)\citenamefont{Isacsson, Gorelik,
  Shekhter, Galperin, and Jonson}}]{isacsson:277002}
\bibinfo{author}{\bibfnamefont{A.}~\bibnamefont{Isacsson}},
  \bibinfo{author}{\bibfnamefont{L.~Y.} \bibnamefont{Gorelik}},
  \bibinfo{author}{\bibfnamefont{R.~I.} \bibnamefont{Shekhter}},
  \bibinfo{author}{\bibfnamefont{Y.~M.} \bibnamefont{Galperin}},
  \bibnamefont{and} \bibinfo{author}{\bibfnamefont{M.}~\bibnamefont{Jonson}},
  \bibinfo{journal}{Phys. Rev. Lett.} \textbf{\bibinfo{volume}{89}},
  \bibinfo{pages}{277002} (\bibinfo{year}{2002}).

\bibitem[{\citenamefont{Fujiwara et~al.}(2004)\citenamefont{Fujiwara,
  Zimmerman, Ono, and Takahashi}}]{Si}
\bibinfo{author}{\bibfnamefont{A.}~\bibnamefont{Fujiwara}},
  \bibinfo{author}{\bibfnamefont{N.~M.} \bibnamefont{Zimmerman}},
  \bibinfo{author}{\bibfnamefont{Y.}~\bibnamefont{Ono}}, \bibnamefont{and}
  \bibinfo{author}{\bibfnamefont{Y.}~\bibnamefont{Takahashi}},
  \bibinfo{journal}{Appl. Phys. Lett.} \textbf{\bibinfo{volume}{84}},
  \bibinfo{pages}{1323} (\bibinfo{year}{2004}).

\bibitem[{\citenamefont{Shilton et~al.}(1996)\citenamefont{Shilton, Talyanskii,
  Pepper, Ritchie, Frost, Ford, Smith, and Jones}}]{saw}
\bibinfo{author}{\bibfnamefont{J.~M.} \bibnamefont{Shilton}},
  \bibinfo{author}{\bibfnamefont{V.~I.} \bibnamefont{Talyanskii}},
  \bibinfo{author}{\bibfnamefont{M.}~\bibnamefont{Pepper}},
  \bibinfo{author}{\bibfnamefont{D.~A.} \bibnamefont{Ritchie}},
  \bibinfo{author}{\bibfnamefont{J.~E.~F.} \bibnamefont{Frost}},
  \bibinfo{author}{\bibfnamefont{C.~J.~B.} \bibnamefont{Ford}},
  \bibinfo{author}{\bibfnamefont{C.~G.} \bibnamefont{Smith}}, \bibnamefont{and}
  \bibinfo{author}{\bibfnamefont{G.~A.~C.} \bibnamefont{Jones}},
  \bibinfo{journal}{J. Phys.: Condens. Matter} \textbf{\bibinfo{volume}{8}},
  \bibinfo{pages}{L531} (\bibinfo{year}{1996}).

\bibitem[{\citenamefont{Niskanen et~al.}(2005)\citenamefont{Niskanen, Kivioja,
  Sepp\"a, and Pekola}}]{Niskanen2005}
\bibinfo{author}{\bibfnamefont{A.~O.} \bibnamefont{Niskanen}},
  \bibinfo{author}{\bibfnamefont{J.~M.} \bibnamefont{Kivioja}},
  \bibinfo{author}{\bibfnamefont{H.}~\bibnamefont{Sepp\"a}}, \bibnamefont{and}
  \bibinfo{author}{\bibfnamefont{J.~P.} \bibnamefont{Pekola}},
  \bibinfo{journal}{Phys. Rev. B} \textbf{\bibinfo{volume}{71}},
  \bibinfo{pages}{012513} (\bibinfo{year}{2005}).

\bibitem[{\citenamefont{Pekola et~al.}(1994)\citenamefont{Pekola, Hirvi,
  Kauppinen, and Paalanen}}]{cbt}
\bibinfo{author}{\bibfnamefont{J.~P.} \bibnamefont{Pekola}},
  \bibinfo{author}{\bibfnamefont{K.~P.} \bibnamefont{Hirvi}},
  \bibinfo{author}{\bibfnamefont{J.~P.} \bibnamefont{Kauppinen}},
  \bibnamefont{and} \bibinfo{author}{\bibfnamefont{M.~A.}
  \bibnamefont{Paalanen}}, \bibinfo{journal}{Phys. Rev. Lett.}
  \textbf{\bibinfo{volume}{73}}, \bibinfo{pages}{2903} (\bibinfo{year}{1994}).

\bibitem[{\citenamefont{Naaman and Aumentado}(2006)}]{Naaman}
\bibinfo{author}{\bibfnamefont{O.}~\bibnamefont{Naaman}} \bibnamefont{and}
  \bibinfo{author}{\bibfnamefont{J.}~\bibnamefont{Aumentado}},
  \bibinfo{journal}{Phys. Rev. B} \textbf{\bibinfo{volume}{73}},
  \bibinfo{pages}{172504} (\bibinfo{year}{2006}).

\bibitem[{\citenamefont{Ingold and Nazarov}(1992)}]{ingold}
\bibinfo{author}{\bibfnamefont{G.-L.} \bibnamefont{Ingold}} \bibnamefont{and}
  \bibinfo{author}{\bibfnamefont{Y.~V.} \bibnamefont{Nazarov}},
  \emph{\bibinfo{title}{{\rm in} Single Charge Tunneling}}, edited by H.~Grabert and M.~H.~Devoret
  (\bibinfo{publisher}{Plenum Press}, \bibinfo{address}{New York}, \bibinfo{year}{1992}), pp.
  \bibinfo{pages}{21--106}.

\bibitem[{\citenamefont{Lotkhov et~al.}(2004)\citenamefont{Lotkhov,
  Bogoslovsky, Zorin, and Niemeyer}}]{lotkhov2004}
\bibinfo{author}{\bibfnamefont{S.~V.} \bibnamefont{Lotkhov}},
  \bibinfo{author}{\bibfnamefont{S.~A.} \bibnamefont{Bogoslovsky}},
  \bibinfo{author}{\bibfnamefont{A.~B.} \bibnamefont{Zorin}}, \bibnamefont{and}
  \bibinfo{author}{\bibfnamefont{J.}~\bibnamefont{Niemeyer}},
  \bibinfo{journal}{J. Appl. Phys.} \textbf{\bibinfo{volume}{95}},
  \bibinfo{pages}{6325} (\bibinfo{year}{2004}).

\bibitem[{\citenamefont{Lotkhov et~al.}(2001)\citenamefont{Lotkhov,
  Bogoslovsky, Zorin, and Niemeyer}}]{Lotkhov}
\bibinfo{author}{\bibfnamefont{S.~V.} \bibnamefont{Lotkhov}},
  \bibinfo{author}{\bibfnamefont{S.~A.} \bibnamefont{Bogoslovsky}},
  \bibinfo{author}{\bibfnamefont{A.~B.} \bibnamefont{Zorin}}, \bibnamefont{and}
  \bibinfo{author}{\bibfnamefont{J.}~\bibnamefont{Niemeyer}},
  \bibinfo{journal}{Appl. Phys. Lett.} \textbf{\bibinfo{volume}{78}},
  \bibinfo{pages}{946} (\bibinfo{year}{2001}).

\bibitem[{\citenamefont{Cholascinski and Chhajlany}(2006)}]{Cholascinski}
\bibinfo{author}{\bibfnamefont{M.}~\bibnamefont{Cholascinski}}
  \bibnamefont{and} \bibinfo{author}{\bibfnamefont{R.~W.}
  \bibnamefont{Chhajlany}} (\bibinfo{year}{2006}), \eprint{cond-mat/0607416
  (unpublished)}.

\bibitem[{\citenamefont{M{\"o}tt{\"o}nen
  et~al.}(2006)\citenamefont{M{\"o}tt{\"o}nen, Pekola, Vartiainen, Brosco, and
  Hekking}}]{Mottonen2006}
\bibinfo{author}{\bibfnamefont{M.}~\bibnamefont{M{\"o}tt{\"o}nen}},
  \bibinfo{author}{\bibfnamefont{J.~P.} \bibnamefont{Pekola}},
  \bibinfo{author}{\bibfnamefont{J.~J.} \bibnamefont{Vartiainen}},
  \bibinfo{author}{\bibfnamefont{V.}~\bibnamefont{Brosco}}, \bibnamefont{and}
  \bibinfo{author}{\bibfnamefont{F.~W.~J.} \bibnamefont{Hekking}},
  \bibinfo{journal}{Phys. Rev. B} \textbf{\bibinfo{volume}{73}},
  \bibinfo{pages}{214523} (\bibinfo{year}{2006}).

\end{thebibliography}
\end{document}